\documentclass[12pt,times]{article}
 \usepackage[pdftex]{color,graphicx}
\usepackage{enumerate}
\usepackage{amsmath}
\usepackage{amssymb}
\usepackage{alltt}
\usepackage{setspace}
\usepackage{subfigure}
\usepackage{sidecap}
\usepackage{sectsty}
\partfont{\large}
\sectionfont{\large}
\subsectionfont{\small}
\usepackage[left=2cm,top=2cm,right=3cm,head=1cm,foot=1cm]{geometry}
\numberwithin{equation}{section}
\let\oldsqrt\sqrt
\def\sqrt{\mathpalette\DHLhksqrt}
\def\DHLhksqrt#1#2{%
\setbox0=\hbox{$#1\oldsqrt{#2\,}$}\dimen0=\ht0
\advance\dimen0-0.2\ht0
\setbox2=\hbox{\vrule height\ht0 depth -\dimen0}%
{\box0\lower0.4pt\box2}}
\newcommand{\al}{\alpha}

\newcommand{\e}{\varepsilon}
\newcommand{\ta}{\theta}

\newcommand{\G}{\Gamma}

\newcommand{\ph}{\varphi}
\newcommand{\da}{\dagger}

\newcommand{\la}{\mathcal L}

\newcommand{\Ld}{\Lambda}
\newcommand{\D}{\mathcal D}
\newcommand{\M}{\mathcal M}

\newcommand{\ml}{\left(\begin{matrix}}
\newcommand{\mr}{\end{matrix}\right)}

\newcommand{\U}{\mathcal U}

\newcommand{\tr}{\text{tr}}

\newcommand{\del}{\delta}

\newcommand{\ka}{\kappa}
\newcommand{\VPMNS}{V_{\text{PMNS}}}
\newcommand{\VCKM}{V_{\text{CKM}}}
\newcommand{\ep}{\epsilon}

\newcommand{\im}{\text{Im}}

\newcommand{\half}{\tfrac{1}{2}}

\newcommand{\fourth}{\tfrac{1}{4}}

\newcommand{\sal}{\!\!\!\sum_{\al\,=\,e,\mu,\tau}\!\!\!}
\newcommand{\delCP}{\del_{\text{CP}}}

\newcommand{\ebar}{\overline e}

\begin{document}
\title{\Large \textbf{The neutrino mixing matrix could (almost) be diagonal with entries $\pm1$}}
\author{Yoni BenTov$^{1}$ \and A. Zee$^{1,2}$}
\date{}
\maketitle
\begin{flushleft}
\textit{$^1$\,Department of Physics, University of California, Santa Barbara CA 93106}\\
\textit{$^2$\,Kavli Institute for Theoretical Physics, University of California, Santa Barbara CA 93106}
\end{flushleft}
\begin{abstract}
It is consistent with the measurement of $\ta_{13} \sim 0.15$ by Daya Bay to suppose that, in addition to being unitary, the neutrino mixing matrix is also almost hermitian, and thereby only a small perturbation from diag$(+1,-1,-1)$ in a suitable basis. We suggest this possibility simply as an easily falsifiable ansatz, as well as to offer a potentially useful means of organizing the experimental data. We explore the phenomenological implications of this ansatz and parametrize one type of deviation from the leading order relation $|V_{e3}| \approx |V_{\tau 1}|$. We also emphasize the group-invariant angle between orthogonal matrices as a means of comparing to data. The discussion is purely phenomenological, without any attempt to derive the condition $V^\da \approx V$ from a fundamental theory.
\end{abstract}
\section{A Phenomenological Ansatz}\label{section:intro}
The neutrino mixing matrix $V$ is defined by $\nu_\al = V_{\al i}\nu_i$, where $\al = e,\mu,\tau$ denotes the charged lepton mass basis (``flavor basis"), and $i = 1,2,3$ denotes the neutrino mass basis. The relevant part of the Lagrangian written in the flavor basis reads
\begin{equation}\label{eq:lagrangian}
\la = -\sal m_\al e_\al \ebar_\al - \tfrac{1}{2}\nu_\al (M_\nu)_{\al\beta}\nu_\beta + h.c.
\end{equation}
In this basis, the neutrino mass matrix is $M_\nu = V^*D_\nu V^\da$, where $D_\nu \equiv \text{diag}(m_1,m_2,m_3)$ with the $m_i$ real and positive.
\\\\ 
Assuming that the three light neutrinos of the Standard Model are Majorana, the magnitudes of the entries of $V$ are constrained by data to be 
\begin{equation}\label{eq:new data}
|V_{\text{exp}}| \approx \ml 0.78\!-\!0.84&0.52\!-\!0.61&0.13\!-\!0.17\\ 0.40\!-\!0.58&0.39\!-\!0.65&0.57\!-\!0.80\\ 0.19\!-\!0.43&0.53\!-\!0.74&0.59\!-\!0.81 \mr\;.
\end{equation}
To obtain Eq.~(\ref{eq:new data}) we have used $0.550 \leq \ta_{12}\leq 0.658$ and $0.620 \leq \ta_{23} \leq 0.934$ from the work of Gonzalez-Garcia, Maltoni, and Salvado \cite{review2}, and $0.135 \leq \ta_{13} \leq 0.171$ from the recent results of Daya Bay \cite{dayabay}. The ranges in Eq.~(\ref{eq:new data}) are correlated in such a way as to preserve the unitarity condition $V^\da V = I$. 
\\\\
In an effort to obtain a theoretical understanding of the mixing matrix, one might suppose that the numerical values in Eq.~(\ref{eq:new data}) arise as a small perturbation from a ``simple" ansatz. As a straw man argument for what such an ansatz might be, consider an older global best fit given by \cite{review}
 \begin{equation}\label{eq:data}
 |V_{\text{exp, old}}| \approx \ml 0.77\!-\!0.86&0.50\!-\!0.63&0.00\!-\!0.22\\0.22\!-\!0.56&0.44\!-\!0.73&0.57\!-\!0.80\\0.21\!-\!0.55&0.40\!-\!0.71&0.59\!-\!0.82 \mr\;.
\end{equation}
Simply by looking at the ranges in Eq.~(\ref{eq:data}), we observe that it was once numerically consistent to suppose that $V$ is hermitian. Without any theoretical motivation, we now suppose that the true mixing matrix satisfies the leading order relation $V^\da \approx V$, and then we study small deviations required to fit the new data. We propose this rather ad hoc constraint in the spirit of trying to make sense of the data by reducing the number of free parameters in the neutrino sector \cite{glashow}. This exercise is intended partially to illustrate that there are still many possibilities for what the true mixing matrix might be.
\section{Real Symmetric Mixing Matrix}\label{section:realsym}
\begin{figure}[h]
\begin{center}
\fbox{
	\begin{minipage}{14 cm}
	\begin{center}
		\includegraphics[scale=1]{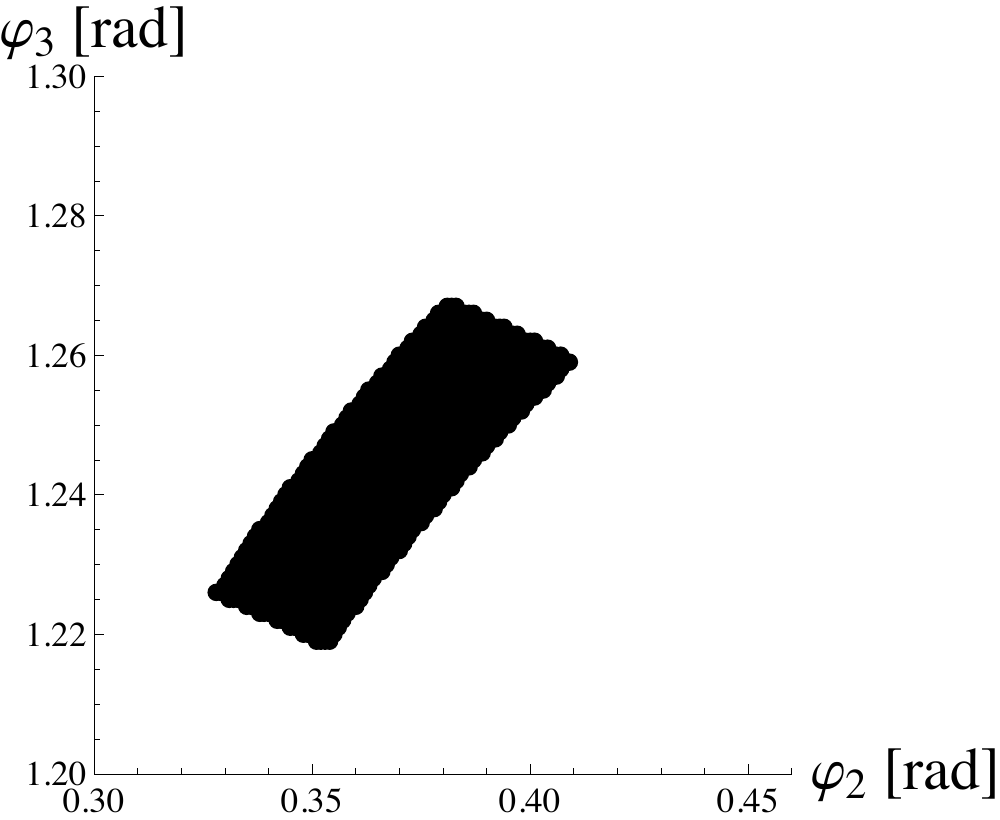}
	\end{center}
	\vspace{-10pt}
	\caption{\small{The values of $\ph_2$ and $\ph_3$ in the quadrant $(\ph_2,\ph_3)\;\ep\;([0,\frac{\pi}{2}],[0,\frac{\pi}{2}])$ consistent with an older set of oscillation data, Eq.~(\ref{eq:data}). This may be a useful starting point about which to perturb in order to fit the new data, Eq.~(\ref{eq:new data}).}}
	\label{fig:angles}	
	\end{minipage}
	}
\end{center}
\end{figure}

As a warmup, we first consider the case for which $V$ is real. Then $V$ is orthogonal, meaning $V^TV = I$, and our hermiticity ansatz amounts to imposing the condition $V^T = V$. The old experimental bounds adjusted for compatibility with this ansatz are
\begin{equation}\label{eq:ansatzdata}
|V_{\text{exp, old}}^{\text{ansatz}}| \approx \ml 0.77\!-\!0.86&0.50\!-\!0.56&0.21\!-\!0.22\\ 0.50\!-\!0.56&0.44\!-\!0.73&0.57\!-\!0.71\\ 0.21\!-\!0.22&0.57\!-\!0.71&0.59\!-\!0.82 \mr\;.
\end{equation}
Again, the ranges are correlated, as required by $V^T V = I$ [See Eq.~(\ref{eq:symmetricV})]. Immediately we see that the Daya Bay observation of $|V_{e3}| < 0.21$ rules out this ansatz as an exact prediction\footnote{Much of this work was completed before the measurement of nonzero reactor angle, $\ta_{13} \sim 0.15$. The fact that the reactor angle is relatively large, meaning closer to $\sim 0.2$ than to zero, is what keeps our analysis relevant.}, but otherwise it is still consistent with Eq.~(\ref{eq:new data}).
\\\\
If $V$ is real symmetric, then it can be diagonalized by an orthogonal transformation: $V = XdX^T$, where $d$ is diagonal and $X$ is orthogonal. Then $V^TV = I$ implies $V^2 = Xd^2X^T = I$, so that $d^2 = I$. Thus our ansatz amounts to proposing that, in a particular basis, the neutrino mixing matrix is diagonal with entries equal to $\pm 1$.
\\\\
We now have a choice as to how to arrange the minus signs in $d$. Two of the nonzero entries in $d$ must have the same sign, while the third must have a sign opposite to that of the first two\footnote{The solution $d = I$ trivially satisfies $d^2 = I$. This would result in $V = XdX^T = XX^T = I$, which is of course experimentally unacceptable.}. In other words, we get to choose which 2-dimensional subspace of $d$ is proportional to the identity matrix. This choice is arbitrary\footnote{For example, let $X = X'P$, where $X'$ is orthogonal and $P$ is a permutation matrix. Then $V \equiv XdX^T = X'd'(X')^T$ is of the same form as before, but with a new diagonal matrix $d' = PdP^T$ with the two minus signs permuted. Of course, we are also free to multiply $V$ and hence $d$ by an overall sign.}; to fix the discussion, we choose
\begin{equation}\label{eq:d}
d = \ml 1&0&0\\0&-1&0\\0&0&-1 \mr
\end{equation}
so that $d$ equals minus the identity matrix in the $(2,3)$-plane.
\\\\
Any rotation matrix in 3 dimensions can be written as a product of independent rotations about each of 3 mutually orthogonal axes. That is, given the rotation matrices
\begin{equation}\label{eq:rotations}
X_1 = \ml 1&0&0\\0&C_1&-S_1\\0&S_1&C_1 \mr\;,\;\;X_2 = \ml C_2&0&S_2\\ 0&1&0\\ -S_2&0&C_2 \mr\;,\;\; X_3 = \ml C_3&S_3&0\\ -S_3&C_3&0\\ 0&0&1 \mr
\end{equation}
where $C_I \equiv \cos\ph_I$ and $S_I \equiv \sin\ph_I$, we can write $X$ as a product of the three $X_I$ in any order\footnote{At this point we should emphasize that $\ph_I$ are not the three PMNS angles that parametrize the mixing matrix $V = XdX^T$. That is why we have chosen to denote their sines and cosines by capital letters, in contrast to the notation in Section \ref{section:hermitian} for the usual PMNS angles.}. Since $d$ is proportional to the identity matrix in the $(2,3)$-plane, it is unchanged by a rotation about the first axis: $X_1 d X_1^T = d$. Thus one of the parameters in $V = XdX^T$ drops out, leaving us with a two-parameter ansatz for the mixing matrix. Choosing the ordering $X = X_3X_2X_1$ implies
\begin{equation}\label{eq:symmetricV}
V = \ml C_2^2\cos(2\ph_3)-S_2^2&-C_2^2\sin(2\ph_3)&-C_3\sin(2\ph_2)\\ \times&-C_2^2\cos(2\ph_3)-S_2^2&S_3\sin(2\ph_2)\\ \times&\times&-\cos(2\ph_2) \mr
\end{equation}
where we have displayed only the upper triangle of $V$ since by construction it is symmetric. The values for the angles consistent with Eq.~(\ref{eq:data}) turn out to be $0.32 \leq \ph_2 \leq 0.42$ and $1.20 \leq \ph_3 \leq 1.27$, where the angles are expressed in radians, as shown in Fig \ref{fig:angles}. 
\\
\\
As an arbitrarily chosen ``typical" example of hermitian mixing, the values $(\ph_2,\ph_3) = (0.35,1.23)$ imply a mixing matrix
\begin{equation}\label{eq:exampleV}
V_{\text{0.35,1.23}} \approx \ml -0.80&0.56&0.22\\ 0.56&0.57&0.61\\ 0.22&0.61&-0.76 \mr\;
\end{equation}
where we have rearranged the minus signs into a standard form\footnote{So as not to interrupt the logical flow we will postpone discussion of rephasing $V$ until Section \ref{section:hermitian}. }. 
Compare this with the often-studied ``tribimaximal mixing" ansatz \cite{wtribi, tribi}
\begin{equation}\label{eq:tribi}
V_{\text{TB}} \equiv \ml \frac{-2}{\sqrt6}&\frac{1}{\sqrt3}&0\\ \frac{1}{\sqrt6}&\frac{1}{\sqrt3}&\frac{1}{\sqrt2}\\ \frac{1}{\sqrt6}&\frac{1}{\sqrt3}&\frac{-1}{\sqrt2} \mr \approx \ml -0.82&0.58&0\\ 0.41&0.58&0.71\\ 0.41&0.58&-0.71 \mr\;.
\end{equation}
These two matrices appear qualitatively ``very different," given that one has $V_{e3} \approx 0.22$ while the other has $V_{e3} = 0$. To make this notion more precise, define\footnote{Another measure \cite{zeeangles} of the $SO(3)$-invariant distance between matrices is $\D(V,V') = \frac{1}{3}\tr(I-V^TV')$. Here we choose the angular distance because it provides an intuitive notion of ``large" versus ``small" in terms of an angle $\Theta$ ranging from 0 to $\pi/2$.} the $SO(3)$-invariant angle $\Theta$ between two special orthogonal matrices $V$ and $V'$:
\begin{equation}
\Theta(V,V') \equiv \cos^{-1}\!\left(\tfrac{1}{3}\tr(V^TV')\right)\;.
\end{equation}
The matrices $V_{0.35,1.23}$ and $V_{\text{TB}}$ are separated by an angle 
\begin{equation}\label{eq:angle between two ansatze}
\Theta(V_{0.35,1.25},V_{\text{TB}}) \approx 0.20 \sim 11^\circ
\end{equation}
in $SO(3)$. As a related example, one might compare to another ansatz with the same atmospheric and reactor angles as tribimaximal mixing $(\ta_{23} = \frac{\pi}{4}$ and $\ta_{13} = 0$, respectively), but with the solar angle related to\footnote{The references discuss various different proposals for relating the solar angle to the golden ratio. We arbitrarily choose the particular implementation of Eq.~(\ref{eq:goldenratio}) to be concrete.} the golden ratio: $\ta_{12} = \tan^{-1}(1/\ph)$, with $\ph = \half(1+\sqrt5)$. \cite{golden ratio 1, golden ratio 2, golden ratio 3, golden ratio 4, golden ratio 5} The PMNS matrix for this ansatz is
\begin{equation}\label{eq:goldenratio}
V_{\text{golden}} \approx \ml -0.85&0.53&0\\ 0.37&0.60&0.71\\ 0.37&0.60&-0.71 \mr\;.
\end{equation}
This is separated from the matrix of Eq.~(\ref{eq:exampleV}) by an angle $\Theta(V_{0.35,1.25},V_{\text{golden}}) \approx 0.20 \sim 11^\circ$, approximately the same as for $V_{\text{TB}}$. More generally, we see that the entire family of ``$\mu\tau$-symmetric" mixing matrices is approximately separated from the entire family of hermitian mixing matrices by $\sim 11^\circ$ in $SO(3)$.
\\\\
We now know, due to Eq.~(\ref{eq:new data}), that neither Eq.~(\ref{eq:exampleV}) nor Eq.~(\ref{eq:tribi}) is correct at low energy, but both may still serve as leading order predictions. The present oscillation data therefore admit two possible starting points that are separated by an angle $\sim 11^\circ$ in the set of all possible 3-by-3 special orthogonal matrices. This is simply intended to give a quantitative measure of uncertainty in our theoretical understanding of the mixing matrix.
\\\\
We might also like to obtain a quantitative sense of how different the new data is from the old. As perhaps an overly simplistic approach \cite{zeeangles}, we extract the arithmetic mean values for the matrices in Eqs.~(\ref{eq:new data}) and~(\ref{eq:data}) and fix the signs according to a chosen convention:
\begin{equation}\label{eq:mean}
V_{\text{exp}}^{\text{mean}} = \ml -0.810&0.565&0.150\\ 0.490&0.520&0.685\\ 0.310&0.635&-0.700 \mr\;,\;\; V_{\text{exp, old}}^{\text{mean}} = \ml -0.815&0.565&0.110\\ 0.390&0.585&0.685\\ 0.380&0.555&-0.705 \mr\;.
\end{equation}
Neither of these matrices is orthogonal. To correct for this, we define ``corrected" versions of these matrices by multiplying on the right by some other matrix $\G$ yet to be determined: $\hat V_{\text{exp}}^{\text{mean}} \equiv V_{\text{exp}}^{\text{mean}}\G$. Then $(\hat V_{\text{exp}}^{\text{mean}})^T\hat V_{\text{exp}}^{\text{mean}} = \G^T(V_{\text{exp}}^{\text{mean}})^TV_{\text{exp}}^{\text{mean}}\G$. The matrix $(V_{\text{exp}}^{\text{mean}})^TV_{\text{exp}}^{\text{mean}}$ is real symmetric, so it can be written as $(V_{\text{exp}}^{\text{mean}})^TV_{\text{exp}}^{\text{mean}} = S\Ld^2 S^T$, where $\Ld^2$ is diagonal with real positive entries, and $S^TS = I$. Thus if we fix\footnote{The ordering of the three eigenvalues in $\Ld^2$ and the corresponding eigenvectors in $S$ is arbitrary and does not change the matrix $\G = S\Ld^{-1}S^T$. This is conceptually the same freedom as that of arranging the signs in $d$ of Eq.~(\ref{eq:d}) and the columns in $X$ without changing $V = XdX^T$.} $\G = S\Ld^{-1}S^T$, we have $(\hat V_{\text{exp}}^{\text{mean}})^T\hat V_{\text{exp}}^{\text{mean}} = I$. 
\\\\
The ``corrected" experimental mixing matrix
\begin{equation}\label{eq:corrected}
\hat V_{\text{exp}}^{\text{mean}} \equiv V_{\text{exp}}^{\text{mean}} S\Ld^{-1}S^T
\end{equation}
is orthogonal. Carrying out this procedure for $V_{\text{exp}}^{\text{mean}}$ and $V_{\text{exp, old}}^{\text{mean}}$ gives
\begin{equation}\label{eq:corrected matrices}
\hat V_{\text{exp}}^{\text{mean}} = \ml -0.811&0.565&0.151\\ 0.494&0.525&0.693\\ 0.312&0.637&-0.705 \mr\;,\;\; \hat V_{\text{exp, old}}^{\text{mean}} = \ml -0.824&0.564&0.053\\ 0.437&0.574&0.693\\ 0.361&0.594&-0.719 \mr\;.
\end{equation}
Using these, we find
\begin{equation}\label{eq:angle between old and new}
\Theta(\hat V_{\text{exp}}^{\text{mean}}, \hat V_{\text{exp, old}}^{\text{mean}}) \approx 0.08 \sim 5^\circ
\end{equation}
\\
Thus, in some quantitative sense, the old data is ``close" to the new data and therefore may help characterize possible ansatze away from which the true mixing matrix might be only a small perturbation. 
\section{Complex Hermitian Mixing Matrix}\label{section:hermitian}
We are now ready to consider a fully complex mixing matrix $V$. If $V$ is hermitian, then it can be diagonalized by a unitary transformation: $V = XdX^\da$, where $d$ is diagonal and $X$ is unitary. The unitarity condition $V^\da V = I$ and the hermiticity ansatz $V^\da = V$ imply $d^2 = I$, just as for the real case. Again we take $d = \text{diag}(+1,-1,-1)$. 
\\\\
The most general unitary 3-by-3 matrix has $3^2 = 9$ independent real parameters. We now briefly recapitulate the justification behind the standard angular parameterization\footnote{Different parametrizations of the neutrino mixing matrix have been studied in the literature \cite{other, other1, other2}.} of a unitary matrix \cite{angles1,angles2}. If $U$ is a unitary matrix, then a matrix $V$ whose elements are $V_{ij} = e^{\,i(x_i+y_j)}U_{ij}$ is also unitary. The angles $x_i$ and $y_j$ together constitute 5 independent parameters, not 6, since they enter only in the combination $x_i+y_j$. Three of the four remaining parameters in $U$ can be taken as the three independent rotations from Section \ref{section:realsym}. The final parameter can be included as a non-removable phase in one of the rotation matrices. 
\\\\
Thus we arrive at the usual PMNS parameterization of the neutrino mixing matrix, $V = \mathcal K \VPMNS \M$, where $\mathcal K \equiv \text{diag}(e^{\,i\ka_1},e^{\,i\ka_2},e^{\,i\ka_3})$, $\M \equiv \text{diag}(e^{\,i\rho},e^{\,i\sigma},1)$ and 
\begin{equation}\label{eq:PMNSangles}
\VPMNS \equiv \ml 1&0&0\\0&c_{23}&s_{23}\\0&s_{23}&-c_{23} \mr\ml c_{13}&0&s_{13}\,e^{-i\del_{\text{CP}}}\\ 0&1&0\\ -s_{13}\,e^{+i\del_{\text{CP}}}&0&c_{13} \mr\ml -c_{12}&s_{12}&0\\ s_{12}&c_{12}&0\\ 0&0&1 \mr
\end{equation}
where $c_{IJ} \equiv \cos\ta_{IJ}$ and $s_{IJ} \equiv \sin\ta_{IJ}$. The matrix $\mathcal K$ is unphysical and can be chosen arbitrarily. The ``Majorana" matrix $\M$ is physical if the neutrinos are Majorana, but it drops out of oscillation probabilities and hence is not observable in oscillation experiments. The four parameters $\ta_{IJ}$ and $\del_{\text{CP}}$ contribute to oscillations. With the chosen sign conventions, the PMNS matrix is
\begin{equation}\label{eq:PMNS}
\VPMNS = \ml -c_{13}c_{12}&c_{13}s_{12}&\hat s_{13}^*\\ c_{23}s_{12}+s_{23}\hat s_{13}c_{12}&c_{23}c_{12}-s_{23}\hat s_{13}s_{12}&s_{23}c_{13}\\ s_{23}s_{12}-c_{23}\hat s_{13}c_{12} & s_{23}c_{12}+c_{23}\hat s_{13}s_{12} & -c_{23}c_{13} \mr\;,\;\; \hat s_{13} \equiv s_{13}\,e^{\,i\delCP}
\end{equation}
which explains the signs in Eqs. (\ref{eq:exampleV}) and (\ref{eq:tribi}). 
\\\\
As discussed in Section \ref{section:realsym}, we are free to choose the sequence in which the three rotations in Eq.~(\ref{eq:rotations}) occur. Moreover, we are free to insert the phase angle $\delCP$ into any of the three rotation matrices in Eq.~(\ref{eq:rotations}). The PMNS parameterization (\ref{eq:PMNSangles}, \ref{eq:PMNS}) is useful for neutrino mixing since the relation $V_{e3} = s_{13}\,e^{-i\del_{\text{CP}}}$ makes clear that $\del_{\text{CP}}$ drops out of the mixing matrix if $V_{e3} = 0$.
\\\\
Even if the above parameterization is convenient for $V$, we should determine the parameterization most convenient for the unitary matrix $X$, which we remind the reader is defined by $V = XdX^\da$. 
\\\\
Since we chose $d = \text{diag}(+1,-1,-1)$ to be proportional to the identity matrix in the $(2,3)$-plane, we should put the analog of $\del_{\text{CP}}$ into a rotation about the first axis. We write $X = N^*X_3X_2\tilde X_1 K$, where $N \equiv \text{diag}(-e^{\,ir},+e^{\,is},+1)$, $K \equiv \text{diag}(e^{\,ik_1},e^{\,ik_2},e^{\,ik_3})$, $X_2$ and $X_3$ are given in Eq. (\ref{eq:rotations}), and finally
\begin{equation}\label{eq:complexX1}
\tilde X_1 \equiv \ml 1&0&0\\0&C_1&-S_1\,e^{-i\eta}\\0&S_1\,e^{+i\eta}&C_1 \mr\;.
\end{equation}
Then both $K$ and $\tilde X_1$ drop out of the mixing matrix $V = XdX^\da$. Putting this into the form $V = \mathcal K\VPMNS\M$ gives\footnote{More explicitly, the matrix $N$ can be written as $N = -d\M$, where $d$ is the matrix of Eq. (\ref{eq:d}) and $\M$ is the matrix of Majorana phases defined above Eq. (\ref{eq:PMNSangles}), so that $r = \rho$ and $s = \sigma$. Then $V = XdX^\da = \M^*(d X_3X_2dX_2^TX_3^Td)\M$ is put into the form $V = \mathcal K\VPMNS\M$ with $\VPMNS = d X_3X_2dX_2^TX_3^Td$ and $\mathcal K = \M^*$.}
\begin{equation}\label{eq:symmetricVPMNS}
\VPMNS = \ml C_2^2\cos(2\ph_3)-S_2^2&C_2^2\sin(2\ph_3)&C_3\sin(2\ph_2)\\ \times&-C_2^2\cos(2\ph_3)-S_2^2&\sin(2\ph_2)S_3\\ \times&\times&-\cos(2\ph_2) \mr
\end{equation}
which is precisely the same PMNS matrix as for the real symmetric mixing matrix ansatz in Eq. (\ref{eq:symmetricV}) up to rearranging the signs. When verifying that the signs are correct, keep in mind that $\sin(2\ph_3) > 0$ but $\cos(2\ph_3) < 0$, and $|\cos(2\ph_3)| > \sin(2\ph_3)$ for the allowed values of $\ph_3$ (see Fig. \ref{fig:angles}).
\\\\
Thus the ansatz of hermiticity $V^\da = V$, if it were exactly true, would predict that neutrino oscillations conserve CP.
\section{Deviations from Hermiticity}\label{section:deviations}
As can be seen from the data in Eq. (\ref{eq:data}), the measurement of $|V_{e3}| < 0.21$ has immediately ruled out the possibility that $V$ is exactly hermitian. Moreover, as shown in Eq. (\ref{eq:symmetricVPMNS}), measuring any CP violation in neutrino oscillations would also signal deviations from $V^\da \approx V$.
\\\\
In either case, $V$ could still be approximately hermitian up to small corrections, so we should find a way to write deviations from hermiticity as a perturbation expansion in a small parameter. In any such parameterization away from $V^\da = V$, we obviously still need to maintain the unitarity condition $V^\da V = I$. 
\\\\
One possibility is to consider deviations from $d^\da = d$ while maintaining $d^\da d = I$. Consider the orthogonal matrix $V_\e \equiv (X_3X_2)d_\e (X_3X_2)^T$, where
\begin{equation}\label{eq:real deviation}
d_\e \equiv \ml 1&0&0\\0&-\cos\e&+\sin\e\\ 0&-\sin\e&-\cos\e \mr\;.
\end{equation}
In the limit $\e \to 0$, this matrix recovers the most general real, orthogonal, symmetric 3-by-3 matrix given in Eqs. (\ref{eq:symmetricV}) and (\ref{eq:symmetricVPMNS}). We have $V = V(\e = 0)+\e\del V+O(\e^2)$, where\footnote{Here we find it convenient to keep the signs as in Eq. (\ref{eq:symmetricV}), and then to convert to the signs in Eq. (\ref{eq:symmetricVPMNS}) only after adding $V(\e = 0)+\e \del V$.}
\begin{equation}\label{eq:deltaV}
\del V = \ml 0&-S_2&+C_2S_3\\ +S_2&0&+C_2C_3\\ -C_2S_3&-C_2C_3&0 \mr
\end{equation}
manifestly parameterizes a particular type of deviation from the real symmetric ansatz. In particular, it parameterizes a decrease in $|V_{e3}|$ and a corresponding increase in $|V_{\tau 1}|$. 
\\\\
Now we are ready to confront the new data of Eq.~(\ref{eq:new data}). The hermitian ansatz (\ref{eq:symmetricVPMNS}) is not compatible with $|V_{\text{exp}}|$ due to the tightened bounds on $V_{e3}$. As a concrete example, recall the unperturbed matrix of (\ref{eq:exampleV}) with $(\ph_2,\ph_3) = (0.35,1.23)$, which we repeat for convenience:
\begin{equation}
V_{0.35,1.23}(\e = 0) \approx \ml -0.80&0.56&0.22\\ 0.56&0.57&0.61\\ 0.22&0.61&-0.76 \mr\;.
\end{equation}
Perturbing this matrix in the form $V = V(\e = 0)+\e\del V$ with $\e = 0.074$ and $\del V$ given in Eq. (\ref{eq:deltaV}) implies a mixing matrix
\begin{equation}
V \approx \ml -0.80&0.58&0.15\\ 0.53&0.57&0.63\\ 0.28&0.58&-0.76 \mr\;
\end{equation}
which is compatible with the bounds in $|V_{\text{exp}}|$.
\\\\
Since we are breaking the hermiticity ansatz, we re-introduce the possibility of CP violation. Thus we can generalize Eq.~(\ref{eq:real deviation}) to the case
\begin{equation}\label{eq:complex deviation}
d_{\e,\del} \equiv \ml 1&0&0\\0&-\cos\e&+e^{+i\del}\sin\e\\ 0&-e^{-i\del}\sin\e&-\cos\e \mr \;,
\end{equation}
which implies a CP angle
\begin{equation}\label{eq:delCPresult}
\delCP = \tan^{-1}\!\left(\frac{S_3\sin\del\sin\e}{(1+\cos\e)C_3S_2+S_3\cos\del\sin\e}\right) = \e\;\frac{\tan(\ph_3)\,\sin\del}{2S_2}+O(\e^2)\;.
\end{equation}
For the previous case $(\ph_2,\ph_3,\e) = (0.35,1.23, 0.074)$ this gives $\del_{\text{CP}} \approx 0.30\sin\del$. The basis-independent Jarlskog invariant $J \equiv \im(V_{e1}V_{e2}^*V_{\mu1}^*V_{\mu2})$ is given to leading order in $\e$ by
\begin{equation}\label{eq:Jresult}
J = -\fourth C_2\sin(4\ph_2)\sin(4\ph_3)\,\e\,\sin\del\,+O(\e^2)
\end{equation}
where the coefficient of $\e\sin\del$ ranges from $0.217$ to $0.229$ over the range of values for $(\ph_2,\ph_3)$ given in Fig.~\ref{fig:angles}.
\\\\
The purpose here is simply to illustrate that although present data falsify the ansatz $V^\da = V$, it is entirely possible that the true mixing matrix is only a small perturbation away from being hermitian. Just as the CKM matrix is a small perturbation from the identity, the neutrino mixing matrix, in a suitable basis, is numerically a small perturbation from a diagonal real matrix whose nonzero entries are $\pm1$.

\section{Discussion}\label{section:end}
We have observed, somewhat as a straw man argument, that at present it is numerically consistent to suppose that the neutrino mixing matrix is almost hermitian. 
\\\\
If the neutrino matrix were hermitian, then it would be related by a change of basis to the matrix $d = \text{diag}(+1,-1,-1)$. Moreover, despite having a large $|V_{e3}|$, a hermitian mixing matrix would necessarily conserve CP in oscillations [Eq.~(\ref{eq:symmetricVPMNS})]. 
\\\\
Since tightening the upper bound on $V_{e3}$ falsifies the hermitian ansatz, we have also presented a simple 1-parameter real parameterization of deviations from hermiticity [Eq.~(\ref{eq:real deviation})], which is useful for classifying perturbations away from the relation $V_{e3} = V_{\tau 1}$. It is also easy to include the CP-violating PMNS angle in this parameterization [Eqs.~(\ref{eq:complex deviation}) and~(\ref{eq:delCPresult})].
\\\\
The lesson here is not whether the condition $V^\da = V$ is approximately true, but rather that it is still experimentally consistent to perturb around an ansatz that is significantly different from the often-studied tribimaximal mixing matrix. We emphasize that one quantitative measure of whether an ansatz is a good starting point is the $SO(3)$-invariant angle from a suitably extracted ``snapshot" of the data [Eqs.~(\ref{eq:angle between two ansatze}) and~(\ref{eq:angle between old and new})].
\\\\
Finally, we invite the reader to find a possible theoretical origin for the condition $V^\da \approx V$. 
\\\\
\textit{Note added: } After our work was submitted, W. Rodejohann called to our attention an earlier study \cite{symmetric} of the condition $|V|^T = |V|$. This also brought to our attention a paper by Joshipura and Smirnov \cite{quarklepton}, which points out that the condition $|V|^T = |V|$ can be obtained in models for which $\U^\da M_{u,d,\ell} \,\U^* = D_{u,d,\ell}$ and $\U^TM_\nu\, \U = D_\nu$, in a self-evident notation, so that to leading order $|\VCKM| = I$ and $|\VPMNS| = |\U^T\U|$.
\\\\
\textit{Acknowledgments:}
\\\\
Part of this work was done in March, 2011 while we were visiting the Academia Sinica in Taipei, Republic of China, whose warm hospitality is greatly appreciated. This research was supported by the NSF under Grant No. PHY07-57035.

\end{document}